\documentclass[12pt] {article}
\usepackage{psfrag}
\usepackage{graphicx}
\usepackage{latexsym,amsfonts}
\usepackage{amssymb}
\begin{document}
\setcounter{page}{1}
\def\theequation{\arabic{section}.\arabic{equation}}
\def\theequation{\thesection.\arabic{equation}}
\setcounter{section}{0}

\title{On the $\Sigma^0 \to \Lambda^0 + \gamma$ decay\\
  in the Effective quark model with chiral $U(3)\times U(3)$ symmetry}

\author{A. N. Ivanov$^{1,2,3}$\,\thanks{Corresponding author. E--mail:
    ivanov@kph.tuwien.ac.at, Tel.: +43--1--58801--14261, Fax:
    +43--1--58801--14299}~\thanks{Permanent Address: State Polytechnic
    University, Department of Nuclear Physics, 195251 St. Petersburg,
    Russian Federation}, A. Ya. Berdnikov$^3$, Ya. A. Berdnikov$^3$, M. Faber$^1$,\\
  V. A. Ivanova$^3$, V. F. Kosmach$^3$, A. V. Nikitchenko$^3$, N. I.
  Troitskaya$^3$}

\date{\today}

\maketitle

\vspace{-0.5in}

\begin{center} {\it $^1$Atominstitut der \"Osterreichischen
    Universit\"aten, Technische Universit\"at Wien, \\ Wiedner
    Hauptstrasse 8-10, A-1040 Wien, \"Osterreich \\ and\\ $^2$Stefan
    Meyer Institut f\"ur subatomare Physik,
    \"Osterreichische Akademie der Wissenschaften,
    Boltzmanngasse 3, A-1090, Wien, \"Osterreich\\ and\\ $^3$State
    Polytechnic
    University, Polytechnicheskaya 29,\\
    195251, St Petersburg, Russian Federation}
\end{center}

\begin{center}
\begin{abstract}
  We analyse the angular distributions of the $\Sigma^0 \to
  \Lambda^0\gamma$ decay rate in the laboratory and in the rest frame
  of the $\Sigma^0$\,--\,hyperon in the dependence on baryon
  polarizations. We calculate the dynamical polarization vector of the
  $\Lambda^0$\,--\,hyperon. Within the Effective quark model with
  chiral $U(3)\times U(3)$ symmetry (PRC {\bf 59}, 451 (1999)) we
  calculate the transition magnetic moment $\mu_{\Sigma^0\Lambda^0}$.
  The theoretical value $\mu_{\Sigma^0\Lambda^0} = -\,1.62$, measured
  in nuclear magnetons, agrees well with the experimental data $\vert
  \mu^{\exp}_{\Sigma^0\Lambda^0}\vert = (1.61 \pm 0.08)$ and the
  theoretical result, predicted within the naive quark model
  $\mu_{\Sigma^0\Lambda^0} = (\sqrt{3}/4)(\mu_{\Sigma^-} -
  \mu_{\Sigma^+}) = (-\,1.57 \pm 0.01)$.
\end{abstract}
\end{center} 


\newpage

\section{Introduction}
\setcounter{equation}{0}

The Effective quark model with chiral $U(3)\times U(3)$ symmetry
\cite{AI1} has been successfully applied to the calculation of the
non--leptonic decays of charmed $\Lambda^+_c$\,--\,baryon \cite{AI2,AI3},
the S--wave amplitudes of $\bar{K}N$ and $K^-d$ scattering at
threshold \cite{IV3,IV4} and the analysis of the polarization
properties of the $\Lambda^0$\,--\,hyperon, produced from the Quark--gluon
plasma \cite{AI4}. The value of the $\sigma_{\pi N}$\,--\,term,
$\sigma_{\pi N} = 60\,{\rm MeV}$ calculated within the Effective quark
model with chiral $U(3)\times U(3)$ symmetry \cite{AI1}, agrees well
with the experimental value $\sigma^{\exp}_{\pi N} =
61^{+\,1}_{-\,4}\,{\rm MeV}$, extracted from the experimental data on
the energy level displacement of the ground state of pionic hydrogen
\cite{PSI2}.

In this paper we apply the Effective quark model with chiral
$U(3)\times U(3)$ symmetry to the calculation of the transition
magnetic moment $\mu_{\Sigma^0\Lambda^0}$, responsible for the
$\Sigma^0 \to \Lambda^0 + \gamma$ decay. A new interest to this decay
is connected with the experimental search for the Kaonic Nuclear
Cluster $K^-pp$ state (the KNC ${^2_{\!\bar{K}}}{\rm H}$) \cite{EDB},
predicted within the Potential model approach by Akaishi and Yamazaki
\cite{TDB1} and the Phenomenological quantum field theory model in
\cite{AI05}.  Experimentally the existence of the KNC
${^2_{\!\bar{K}}}{\rm H}$ is analysed by means of the invariant--mass
spectrum of the $\Lambda^0p$ pair, induced by the decays
${^2_{\!\bar{K}}}{\rm H} \to \Lambda^0 + p$ and ${^2_{\!\bar{K}}}{\rm
  H} \to \Sigma^0 + p$. Due to the decay $\Sigma^0 \to \Lambda^0 +
\gamma$ the experimental invariant--mass spectrum of the $\Lambda^0p$
pair should contain two peaks. One of these peaks is located at the
mass of the KNC ${^2_{\!\bar{K}}}{\rm H}$. It is defined by the
${^2_{\!\bar{K}}}{\rm H} \to \Lambda^0 + p$ decay. Another one is
shifted to left from the mass of the KNC ${^2_{\!\bar{K}}}{\rm H}$ by
about $78\,{\rm MeV}$. This shift is defined by the photon energy in
the ${^2_{\!\bar{K}}}{\rm H} \to \Sigma^0 + p \to \Lambda^0 + \gamma +
p$ cascade decay.

The paper is organized as follows. In Section 2 we calculate the
angular distributions of the $\Sigma^0 \to \Lambda^0\gamma$ decay rate
in the laboratory and in the rest frame of the $\Sigma^0$\,--\,hyperon
in dependence on the polarizations of the baryons.  We calculate the
dynamical polarization vector of the $\Lambda^0$\,--\,hyperon. In
Section 3 we calculate the transition magnetic moment
$\mu_{\Sigma^0\Lambda^0} = -\,1.62$, measured in nuclear magnetons.
The calculation is carried out in the Effective quark model with
chiral $U(3)\times U(3)$ symmetry. The theoretical result agrees well
with the experimental value $\vert
\mu^{(\exp)}_{\Sigma^0\Lambda^0}\vert = (1.61 \pm 0.08)$ \cite{PDG04}
and the theoretical prediction within the naive quark model
$\mu_{\Sigma^0\Lambda^0} = -\,1.57 \pm 0.01$ \cite{PDG04a}. In the
Conclusion we discuss the obtained results.

\section{Angular distribution of the $\Sigma^0 \to \Lambda^0 \gamma$
  decay rate}

The $\Sigma^0 \to \Lambda^0 \gamma$ decay is the $M1$ electromagnetic
transition. The amplitude of this transition is defined by
\begin{eqnarray}\label{label2.1}
M(\Sigma^0 \to \Lambda^0 \gamma) = \sqrt{4\pi
  \alpha}\,e^*_{\mu}(k,\lambda)\langle \Lambda^0(k_{\Sigma^0},\sigma_{\Sigma^0})\vert
J^{\mu}(0)\vert \Sigma^0(k_{\Lambda^0},\sigma_{\Lambda^0})\rangle,
\end{eqnarray}
where $J^{\mu}(0)$ is the electromagnetic hadronic current,
$e^*_{\mu}(k,\lambda)$ is a polarization vector of a photon and
$\alpha = 1/137.036$ is the fine--structure constant.  Due to the $M1$
transition the matrix element of the electromagnetic hadronic current
is
\begin{eqnarray}\label{label2.2}
\langle \Lambda^0(k_{\Sigma^0},\sigma_{\Sigma^0})\vert J^{\mu}(0)\vert
\Sigma^0(k_{\Lambda^0},\sigma_{\Lambda^0})\rangle = -\,g_{\Sigma^0\Lambda^0}\,
\bar{u}_{\Lambda^0}(k_{\Lambda^0},\sigma_{\Lambda^0})\sigma^{\mu\nu} k_{\nu} 
u_{\Sigma^0}(k_{\Sigma^0},\sigma_{\Sigma^0}),
\end{eqnarray}
where $\sigma^{\mu\nu} = (\gamma^{\mu}\gamma^{\nu} -
\gamma^{\nu}\gamma^{\mu})/2$, $g_{\Sigma^0\Lambda^0}$ is the effective
coupling constant of the transition $\Sigma^0 \to \Lambda^0$, caused
by strong low--energy interactions, and $k = k_{\Sigma^0} -
k_{\Lambda^0}$, and
$\bar{u}_{\Lambda^0}(k_{\Lambda^0},\sigma_{\Lambda^0})$ and
$u_{\Sigma^0}(k_{\Sigma^0},\sigma_{\Sigma^0})$ are the bispinors of
the $\Lambda^0$ and $\Sigma^0$ hyperons, normalized by
$\bar{u}_Y(k_Y,\sigma_Y)u_Y(k_Y,\sigma_Y) = 2m_Y$ for $Y = \Sigma^0$
and $\Lambda^0$.

The width of the $\Sigma^0 \to \Lambda^0 \gamma$ decay for unpolarized
baryons is proportional to
\begin{eqnarray}\label{label2.3}
  \hspace{-0.3in} &&\frac{1}{2}\sum_{\sigma_{\Lambda^0} = \pm 1/2}
\sum_{\sigma_{\Sigma^0} = \pm
    1/2}\sum_{\lambda = \pm 1}\vert M(\Sigma^0 \to \Lambda^0
  \gamma)\vert^2 =\nonumber\\
  \hspace{-0.3in}  &&= 2\pi\alpha g^2_{\Sigma^0\Lambda^0}\sum_{\lambda = \pm 1}
  e^*_{\mu}(k,\lambda)e_{\alpha}(k,\lambda)k_{\nu}k_{\beta}\,{\rm
    tr}\{(m_{\Sigma^0} + \hat{k}_{\Sigma^0})\sigma^{\mu\nu}(m_{\Lambda^0} +
  \hat{k}_{\Lambda^0})\sigma^{\alpha\beta}\} =\nonumber\\
  \hspace{-0.3in}  &&= 2\pi\alpha g^2_{\Sigma^0\Lambda^0}\sum_{\lambda = \pm 1}
  e^*_{\mu}(k,\lambda)e_{\alpha}(k,\lambda)k_{\nu}k_{\beta}\,{\rm
    tr}\{\hat{k}_{\Sigma^0}\sigma^{\mu\nu} \hat{k}_{\Lambda^0}\sigma^{\alpha\beta}\} =
  32\pi \alpha\, g^2_{\Sigma^0\Lambda^0} (k\cdot k_{\Sigma^0})(k\cdot k_{\Lambda^0}) =
\nonumber\\
  \hspace{-0.3in} &&=  32\pi \alpha\,
  g^2_{\Sigma^0\Lambda^0}\,m^2_{\Sigma^0}\omega^2_0,
\end{eqnarray}
where $\omega_0 = (m^2_{\Sigma^0} - m^2_{\Lambda^0})/2 m_{\Lambda^0} =
74.5\,{\rm MeV}$ \cite{PDG04} is the photon energy in the rest frame
of the $\Sigma^0$\,--\,hyperon.

The width of the $\Sigma^0 \to \Lambda^0 \gamma$ decay is equal to
\begin{eqnarray}\label{label2.4}
  \hspace{-0.3in} \Gamma(\Sigma^0 \to \Lambda^0\gamma) &=& 
  \frac{\alpha}{\pi}\,\frac{g^2_{\Sigma^0\Lambda^0}}{m_{\Sigma^0}}\int
  \delta^{(4)}(k_{\Lambda^0} + k - k_{\Sigma^0})\,(k\cdot k_{\Sigma^0})^2\,
    \frac{d^3k_{\Lambda^0}}{E_{\Lambda^0}(\vec{k}_{\Lambda^0})}
    \frac{d^3k}{\omega(\vec{k})} = \nonumber\\
    &=& 4\alpha g^2_{\Sigma^0\Lambda^0}\,\omega^3_0.
\end{eqnarray}
Using the experimental value for the width of the $\Sigma^0 \to
\Lambda^0\gamma$ decay $\Gamma_{\exp}(\Sigma^0 \to \Lambda^0\gamma)
=(8.9 \pm 0.8)\times 10^{-3}\,{\rm MeV}$ \cite{PDG04} we get
$g_{\Sigma^0\Lambda^0} = (8.6\pm 0.4)\times 10^{-4}\,{\rm MeV}^{-1}$.
The effective coupling constant $g_{\Sigma^0\Lambda^0}$ can be
expressed in terms of the transition magnetic moment
$g_{\Sigma^0\Lambda^0} = \mu_{\Sigma^0\Lambda^0}/2m_N$, where
$\vert \mu_{\Sigma^0\Lambda^0}\vert = 1.61 \pm 0.08$ \cite{PDG04}.

In the laboratory frame, where the $\Sigma^0$\,--\,hyperon moves with
a velocity $\vec{v}_{\Sigma^0} = \vec{k}_{\Sigma^0}/E_{\Sigma^0}$, the
angular distribution of the $\Sigma^0 \to \Lambda^0 \gamma$ decay rate
can be investigated with respect to both the relative angle
$\vartheta_{\Sigma^0\Lambda^0}$ of the $\Sigma^0\Lambda^0$ pair and
the relative angle $\vartheta_{\Sigma^0\gamma}$ of the
$\Sigma^0\gamma$ pair.

For the unpolarized baryons the angular distribution of the
$\Lambda^0$\,--\,hyperon relative to the direction of the motion of the
$\Sigma^0$\,--\,hyperon is
\begin{eqnarray}\label{label2.5}
  4\pi\,\frac{dB(\Sigma^0 \to \Lambda^0\gamma)}{d\Omega_{\Sigma^0\Lambda^0}} = 
  \frac{1}{\gamma_{\Sigma^0}\omega_0}\,\frac{\vec{k}^{\;2}_{\Lambda^0}\,
    \sqrt{1 - \vec{v}^{\;2}_{\Sigma^0}}}{\vert\vec{k}_{\Lambda^0}
    \vert  - E_{\Lambda^0}\vert\vec{v}_{\Sigma^0}\vert\,\cos\vartheta_{\Sigma^0\Lambda^0}},
\end{eqnarray}
where $\gamma_{\Sigma^0} = 1/\sqrt{1 - \vec{v}^{\;2}_{\Sigma^0}}$.
The kinematical parameters of the baryons are related by
$E_{\Lambda^0} - \vert\vec{k}_{\Lambda^0} \vert
\vert\vec{v}_{\Sigma^0}\vert\,\cos\vartheta_{\Sigma^0\Lambda^0} =
(m_{\Sigma^0} - \omega_0)/\gamma_{\Sigma^0}$.  This allows to obtain
the energy spectrum of the $\Lambda^0$\,--\,hyperon
\begin{eqnarray}\label{label2.6}
  \frac{dB(\Sigma^0 \to \Lambda^0\gamma)}{dE_{\Lambda^0}} = 
  \frac{1}{2 \gamma_{\Sigma^0} \vert\vec{v}_{\Sigma^0}\vert\,\omega_0}.
\end{eqnarray}
In turn, the angular distribution of the photon relative to the
direction of the motion of the $\Sigma^0$\,--\,hyperon takes the form
\begin{eqnarray}\label{label2.7}
  4\pi\,\frac{dB(\Sigma^0 \to \Lambda^0\gamma)}{d\Omega_{\Sigma^0\gamma}} = 
  \frac{1}{4\pi \gamma^2_{\Sigma^0}}\,
  \frac{1}{(1 - 
    \vert\vec{v}_{\Sigma^0}\vert\,\cos\vartheta_{\Sigma^0\gamma})^2}.
\end{eqnarray}
The photon energy in the laboratory frame is related to the photon
energy in the center of mass frame as $\omega_0 = \gamma_{\Sigma^0}
\omega\,(1 -
\vert\vec{v}_{\Sigma^0}\vert\,\cos\vartheta_{\Sigma^0\gamma})$. This
allows to calculate the energy spectrum of the photon
\begin{eqnarray}\label{label2.8}
  \frac{dB(\Sigma^0 \to \Lambda^0\gamma)}{d\omega} = 
  \frac{1 }{2 \gamma_{\Sigma^0} \vert\vec{v}_{\Sigma^0}\vert\,\omega_0 } = 
  \frac{1}{\omega_+ - \omega_-},
\end{eqnarray}
where $\omega_+ = \omega_0/\gamma_{\Sigma^0}(1 -
\vert\vec{v}_{\Sigma^0}\vert)$ and $\omega_- =
\omega_0/\gamma_{\Sigma^0}(1 + \vert\vec{v}_{\Sigma^0}\vert)$ are the
maximal and minimal energy of the photon, emitted by the
$\Sigma^0$\,--\,hyperon in the laboratory frame.

In the rest frame of the $\Sigma^0$\,--\,hyperon the angular
distribution of the decay rate is equal to
\begin{eqnarray}\label{label2.9}
  4\pi\,\frac{dB(\Sigma^0 \to \Lambda^0\gamma)}{d\Omega_{\Lambda^0}} = 1,
\end{eqnarray}
where $d\Omega_{\Lambda^0} = 2\pi
\sin\vartheta_{\Lambda^0}d\vartheta_{\Lambda^0}$ and
$\vartheta_{\Lambda^0}$ is polar angle relative to the momentum
$\vec{k}_{\Lambda^0} = -\,\vec{k}$ of the $\Lambda^0$\,--\,hyperon at
the rest frame of the $\Sigma^0$\,--\,hyperon and $\vartheta_{\gamma}
= \pi - \vartheta_{\Lambda^0}$.

For the polarized hyperons, which we denote as $\vec{\Sigma}^0$ and
$\vec{\Lambda}^0$, the expression (\ref{label2.3}) changes as follows
\cite{LL71} (see also \cite{AI01}):
\begin{eqnarray}\label{label2.10}
  &&\sum_{\lambda = \pm 1}\vert M(\vec{\Sigma}^0 \to \vec{\Lambda}^0
  \gamma)\vert^2  =  2\pi\alpha g^2_{\Sigma^0\Lambda^0}\sum_{\lambda = \pm 1}
  e^*_{\mu}(k,\lambda)e_{\alpha}(k,\lambda)k_{\nu}k_{\beta}\nonumber\\
  &&\times\,{\rm
    tr}\{(m_{\Sigma^0} + \hat{k}_{\Sigma^0})(1 + \gamma^5\hat{\zeta}_{\Sigma^0})
  \sigma^{\mu\nu}(m_{\Lambda^0} +
  \hat{k}_{\Lambda^0})(1 + \gamma^5\hat{\zeta}_{\Lambda^0})\sigma^{\alpha\beta}\}.
\end{eqnarray}
Here $\hat{\zeta}_Y = \gamma_{\mu}\zeta^{\mu}_Y$ for $Y = \Sigma^0,
\Lambda^0$, where $\zeta^{\mu}_Y$ is the 4--polarization vector of the
$Y$\,--\,hyperon \cite{LL71} (see also \cite{AI01}):
\begin{eqnarray}\label{label2.11}
  \zeta^{\mu}_Y = \Bigg(\frac{\vec{k}_Y\cdot\vec{\xi}_{\,Y}}{m_Y},\vec{\xi}_{\,Y} +
  \frac{\vec{k}_Y\,(\vec{k}_Y\cdot\vec{\xi}_{\,Y})}{m_Y (E_Y(\vec{k}_Y) +
    m_Y)}\Bigg),
\end{eqnarray}  
where $\vec{\xi}_Y$ is a 3--dimensional polarization vector of the
$Y$\,--\,hyperon, normalized to unity $\vec{\xi}^{\;2}_Y = 1$. The
polarization vector $\zeta^{\mu}_Y$ satisfies the constraints
$\zeta^2_Y = - 1$ and $k_Y\cdot \zeta_Y = 0$.

The result of the calculation of the trace in (\ref{label2.10}) is
\begin{eqnarray}\label{label2.12}
  \hspace{-0.3in}&&\sum_{\lambda = \pm 1}\vert M(\vec{\Sigma}^0 \to \vec{\Lambda}^0
  \gamma)\vert^2 =  32\pi\alpha g^2_{\Sigma^0\Lambda^0}\Big\{(k\cdot k_{\Sigma^0})^2 - 
  (k\cdot k_{\Sigma^0})^2(\zeta_{\Sigma^0}\cdot \zeta_{\Lambda^0}) \nonumber\\
  \hspace{-0.3in}&& - 
  m_{\Sigma^0}(m_{\Sigma^0} + m_{\Lambda^0})\,
  (k\cdot \zeta_{\Sigma^0})(k\cdot \zeta_{\Lambda^0}) -
  \frac{1}{\omega^2}\,
  (k\cdot k_{\Sigma^0})^2\,
  (\vec{k}\times \vec{\zeta}_{\Sigma^0})\cdot 
  (\vec{k}\times \vec{\zeta}_{\Lambda^0})\nonumber\\
  \hspace{-0.3in}&& -\frac{1}{\omega^2}\,(\vec{k}\times \vec{k}_{\Sigma^0})\cdot
  (\vec{k}\times \vec{k}_{\Sigma^0})\,(k\cdot \zeta_{\Sigma^0})\,
  (k\cdot \zeta_{\Lambda^0}) + \frac{1}{\omega^2}\,(k\cdot k_{\Sigma^0})(k\cdot
  \zeta_{\Lambda^0})(\vec{k}
  \times \vec{k}_{\Sigma^0})\cdot 
  (\vec{k}\times \vec{\zeta}_{\Sigma^0})\nonumber\\
  \hspace{-0.3in}&& + \frac{1}{\omega^2}\,(k\cdot k_{\Sigma^0})(k\cdot 
  \zeta_{\Sigma^0})(\vec{k}\times \vec{k}_{\Lambda^0})\cdot 
  (\vec{k}\times \vec{\zeta}_{\Lambda^0})\Big\}.
\end{eqnarray}
We have carried out the summation over the transverse  polarizations of
the photon.

The angular distribution of the $\vec{\Lambda}^0$\,--\,hyperon
relative to the direction of the motion of the
$\vec{\Sigma}^0$\,--\,hyperon can be written in the following general
form \cite{LL71}
\begin{eqnarray}\label{label2.13}
  4\pi \frac{dB(\vec{\Sigma}^0 \to
    \vec{\Lambda}^0 \gamma)}{d\Omega_{\Sigma^0\Lambda^0}} = 
  4\pi \frac{dB(\Sigma^0 \to
    \Lambda^0 \gamma)}{d\Omega_{\Sigma^0\Lambda^0}}\,
(1 + \vec{P}_{\Lambda^0}\cdot \vec{\xi}_{\Lambda^0}),
\end{eqnarray}  
where $\vec{P}_{\Lambda^0}$ is a dynamical polarization vector of the
$\Lambda^0$\,--\,hyperon. Following the procedure expounded in \cite{LL71}
we obtain
\begin{eqnarray}\label{label2.14}
  \hspace{-0.3in}&&\vec{P}_{\Lambda^0} = \Big\{\frac{m_{\Sigma^0} + 
    m_{\Lambda^0}}{m_{\Sigma^0}\omega^2_0}
  (k_{\Lambda^0}\cdot \zeta_{\Sigma^0}) - 
  \frac{(\vec{k}_{\Sigma^0}\times \vec{k}_{\Lambda^0})^2}{\vert \vec{k}_{\Sigma^0} - 
    \vec{k}_{\Lambda^0}\vert^2}\frac{k_{\Lambda^0}\cdot
    \zeta_{\Sigma^0}}{m^2_{\Sigma^0}\omega^2_0} 
  - \frac{(\vec{k}_{\Sigma^0}\times \vec{k}_{\Lambda^0})\cdot 
    ((\vec{k}_{\Sigma^0} - \vec{k}_{\Lambda^0})\times
    \vec{\zeta}_{\Sigma^0})}{\vert \vec{k}_{\Sigma^0} 
    - \vec{k}_{\Lambda^0}\vert^2 m_{\Sigma^0}\omega_0}\Big\}
  \nonumber\\  
  \hspace{-0.3in} &&\times \Big[E_{\Sigma^0}
  \frac{\vec{k}_{\Lambda^0}}{m_{\Lambda^0}} - 
  \vec{k}_{\Sigma^0} - \frac{(\vec{k}_{\Sigma^0}\cdot \vec{k}_{\Lambda^0})
    \vec{k}_{\Lambda^0}}{m_{\Lambda^0}(E_{\Lambda^0} + m_{\Lambda^0})}\Big]- 
  \Big(\frac{\vec{k}_{\Sigma^0}\cdot \vec{\zeta}_{\Sigma^0}}{E_{\Sigma^0}}\Big)
  \frac{\vec{k}_{\Lambda^0}}{m_{\Lambda^0}} + 
  \frac{(\vec{k}_{\Lambda^0}\cdot \vec{\zeta}_{\Sigma^0})
    \vec{k}_{\Lambda^0}}{m_{\Lambda^0}(E_{\Lambda^0} + m_{\Lambda^0})}\nonumber\\
  \hspace{-0.3in}&&+ \frac{1}{\vert \vec{k}_{\Sigma^0} - \vec{k}_{\Lambda^0}\vert^2}
  \Big\{((\vec{k}_{\Sigma^0} - \vec{k}_{\Lambda^0})\cdot \vec{\zeta}_{\Sigma^0}) 
  (\vec{k}_{\Sigma^0} - \vec{k}_{\Lambda^0}) - 
  \frac{(((\vec{k}_{\Sigma^0} - \vec{k}_{\Lambda^0})\times 
    \vec{\zeta}_{\Sigma^0})\cdot (
    \vec{k}_{\Sigma^0}\times \vec{k}_{\Lambda^0}))\vec{k}_{\Lambda^0}}{
    m_{\Lambda^0}(E_{\Lambda^0} + m_{\Lambda^0})}\nonumber\\
  \hspace{-0.3in} && + \frac{k_{\Lambda^0}\cdot 
    \zeta_{\Sigma^0}}{m_{\Sigma^0}\omega_0}\Big[
  (\vec{k}_{\Sigma^0}\times \vec{k}_{\Lambda^0})\times (\vec{k}_{\Sigma^0} - 
  \vec{k}_{\Lambda^0}) + \frac{(\vec{k}_{\Sigma^0}\times 
    \vec{k}_{\Lambda^0})^2 
    \vec{k}_{\Lambda^0}}{m_{\Lambda^0}(E_{\Lambda^0} + m_{\Lambda^0})}\Big]\Big\}.
\end{eqnarray}  
In the rest frame of the $\vec{\Sigma}^0$\,--\,hyperon the dynamical
polarization vector of the $\vec{\Lambda}^0$\,--\,hyperon takes the
form
\begin{eqnarray}\label{label2.15}
  \vec{P}_{\Lambda^0} = - \frac{m_{\Sigma^0}}{m_{\Lambda^0}}\,
(\vec{n}_{\Lambda^0}\cdot \vec{\xi}_{\Sigma^0})\vec{n}_{\Lambda^0},
\end{eqnarray}
where $\vec{n}_{\Lambda^0} = \vec{k}_{\Lambda^0}/\omega_0$. Since in
the rest frame of the $\vec{\Sigma}^0$\,--\,hyperon
$\vec{k}_{\Lambda^0} = -\,\vec{k}$, $\vec{n}_{\Lambda^0}$ is a unit
vector $\vec{n}^{\,2}_{\Lambda^0} = 1$.

The dynamical polarization of the $\vec{\Lambda}^0$\,--\,hyperon,
averaged over all directions of the vector $\vec{n}_{\Lambda^0}$,
amounts to
\begin{eqnarray}\label{label2.16}
  \langle \vec{P}_{\Lambda^0}\rangle  = -\,\frac{1}{3}\,
\frac{m_{\Sigma^0}}{m_{\Lambda^0}}\,\vec{\xi}_{\Sigma^0}. 
\end{eqnarray}
Our results (\ref{label2.15}) and (\ref{label2.16}) agree well with
those obtained in \cite{RG57}--\cite{NB61}.

\section{Transition magnetic moment $\mu_{\Sigma^0\Lambda^0}$}
\setcounter{equation}{0}

The transition magnetic moment $\mu_{\Sigma^0\Lambda^0}$ responsible
for the $\Sigma^0 \to \Lambda^0 + \gamma$ decay is related to the
effective coupling constant $g_{\Sigma^0\Lambda^0}$ by
$\mu_{\Sigma^0\Lambda^0} = 2m_Ng_{\Sigma^0\Lambda^0}$. The calculation
of the effective coupling constant $g_{\Sigma^0\Lambda^0}$ we carry
out in the Effective quark model with chiral $U(3)\times U(3)$
symmetry \cite{AI1}--\cite{AI4}. For this aim we calculate the matrix
element of the electromagnetic hadronic current $\langle
\Lambda^0(k_{\Sigma^0},\sigma_{\Sigma^0})\vert J^{\mu}(0)\vert
\Sigma^0(k_{\Lambda^0},\sigma_{\Lambda^0})\rangle$.

After the application the reduction technique and the equations of
motion \cite{AI1,AI2} the matrix element $\langle
\Lambda^0(k_{\Sigma^0},\sigma_{\Sigma^0})\vert J^{\mu}(0)\vert
\Sigma^0(k_{\Lambda^0},\sigma_{\Lambda^0})\rangle$ takes the form
\begin{eqnarray}\label{label3.1}
  \hspace{-0.3in}&&\langle
  \Lambda^0(k_{\Sigma^0},\sigma_{\Sigma^0})\vert J^{\mu}(0)\vert
  \Sigma^0(k_{\Lambda^0},\sigma_{\Lambda^0})\rangle =  -\,\frac{1}{2}\,g^2_B\int d^4xd^4y
  \,e^{\textstyle\,ik_{\Lambda^0}\cdot x - ik_{\Sigma^0}\cdot y}
 \nonumber\\
  \hspace{-0.3in}&&\hspace{1.5in}\times\, \bar{u}_{\Lambda^0}(k_{\Lambda^0},\sigma_{\Lambda^0})
  \,\langle 0\vert {\rm T}(\eta_{\Lambda^0}(x)J_{\mu}(0)\bar{\eta}_{\Sigma^0}(y))
  \vert 0\rangle_c\, u_{\Sigma^0}(k_{\Sigma^0},\sigma_{\Sigma^0}),
\end{eqnarray}
where ${\rm T}$ is a time--ordering operator, the index $c$ in
$\langle 0\vert \ldots\vert 0\rangle_c$ means the calculation of the
{\it connected} part of the vacuum expectation value, $J_{\mu}(0)$ is
the electromagnetic quark current
\begin{eqnarray}\label{label3.2}
  J_{\mu}(0) = \frac{2}{3}\,\bar{u}_{\ell}(0)\gamma_{\mu}u_{\ell}(0) - 
\frac{1}{3}\,\bar{d}_{\ell}(0)\gamma_{\mu}d_{\ell}(0) 
  - \frac{1}{3}\,\bar{s}_{\ell}(0)\gamma_{\mu}s_{\ell}(0),
\end{eqnarray}
where $u_{\ell}(0)$, $d_{\ell}(0)$ and $s_{\ell}(0)$ are current quark fields. The
summation over the ``colour'' index ${\ell} = 1,2,3$ is assumed. Then,
$\eta_{\Lambda^0}(x)$ and $\bar{\eta}_{\Sigma^0}(y)$ are the
three--quark densities \cite{AI1,AI2,QSR}
\begin{eqnarray}\label{label3.3}
\eta_{\Lambda^0}(x) &=& - \sqrt{\frac{2}{3}}\,\varepsilon^{ijk}\{[\bar{u^c}_i(x)\gamma_{\mu}s_j(x)]
\,\gamma^{\mu}\gamma^5 d_k(x) - [\bar{d^c}_i(x)\gamma_{\mu}s_j(x)]
\,\gamma^{\mu}\gamma^5 u_k(x)\},\nonumber\\
\eta_{\Sigma^0}(x) &=& - \sqrt{\frac{1}{2}}\,\varepsilon^{ijk}\{[\bar{u^c}_i(x)
\gamma_{\mu}d_j(x)]\,\gamma^{\mu}\gamma^5 s_k(x) + [\bar{d^c}_i(x)\gamma_{\mu}u_j(x)]
\,\gamma^{\mu}\gamma^5 s_k(x)\}
\end{eqnarray}
coupled to the $\Lambda^0$ and $\Sigma^0$
hyperons \cite{AI1,AI2}
\begin{eqnarray}\label{label3.4}
{\cal L}_{\rm eff}(x) = \frac{g_B}{\sqrt{2}}\bar{\Lambda}^0(x)\,\eta_{\Lambda^0}(x)  + 
\frac{g_B}{\sqrt{2}}\bar{\Sigma}^0(x)\,\eta_{\Sigma^0}(x) + {\rm h.c.}.
\end{eqnarray}
The three--quark density $\bar{\eta}_{\Sigma^0}(x)$ is equal to
\begin{eqnarray}\label{label3.5}
  \bar{\eta}_{\Sigma^0}(x)= + \sqrt{\frac{1}{2}}\,\varepsilon^{ijk}\{\bar{s}_i(x)
\gamma^{\mu}\gamma^5 
[\bar{d}_j(x)
\gamma_{\mu}u^c_k(x)] + \bar{s}_i(x)\gamma^{\mu}\gamma^5 
[\bar{u}_j(x)\gamma_{\mu}d^c_k(x)]\}.
\end{eqnarray}
The coupling constant $g_B$ is calculated in \cite{AI1,AI1a}: $g_B = 1.37
\times 10^{-4}\,{\rm MeV}^{-2}$.

The matrix element (\ref{label3.1}) is defined by Feynman diagrams in
Fig.1.
\begin{figure}
\centering
\psfrag{x}{$x$}
\psfrag{y}{$y$}
\psfrag{S0}{$\Sigma^0$}
\psfrag{L0}{$\Lambda^0$}
\psfrag{u}{$u$}
\psfrag{d}{$d$}
\psfrag{s}{$s$}
\psfrag{g}{$\gamma$} 
\psfrag{p1}{$k_{\Sigma^0}$}
\psfrag{p2}{$k_{\Sigma^0} + k_1 - k_2$}
\psfrag{p3}{$k_{\Lambda^0}$}
\psfrag{p4}{$k_2$}
\psfrag{p5}{$k_{\Sigma^0} - k_{\Lambda^0} + k_1$}
\psfrag{p6}{$k_1$}
\psfrag{p7}{$k_{\Lambda^0} + k_2 - k_1$}
\includegraphics[height=0.35\textheight]{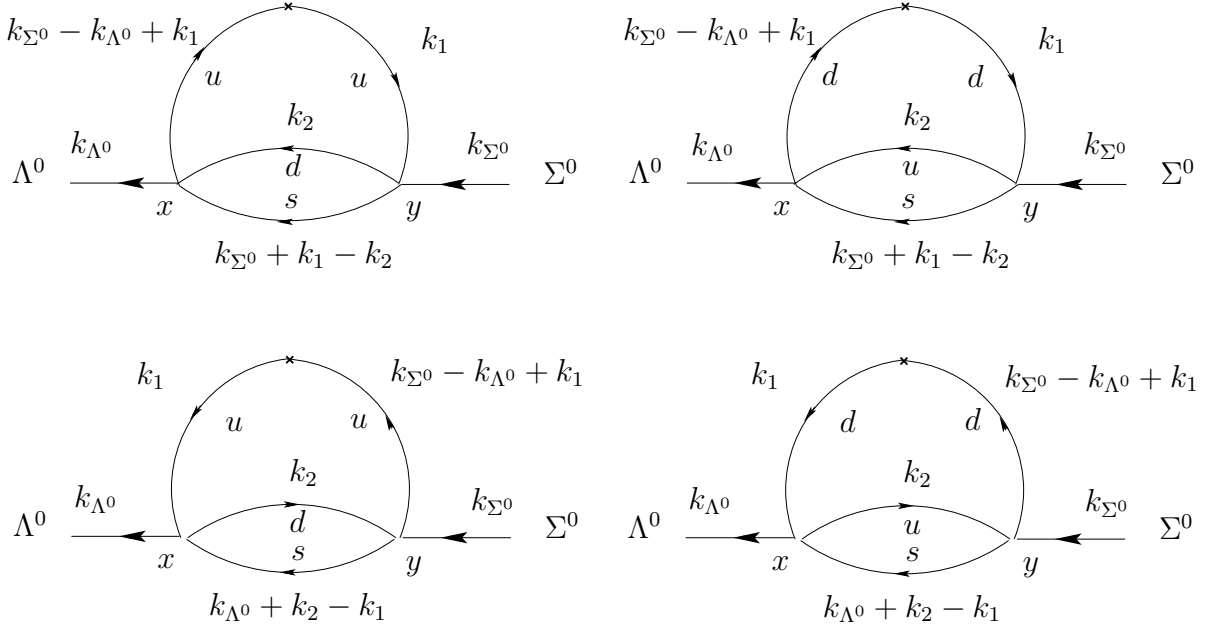}
\caption{Feynman diagrams of the $\Sigma^0 \to \Lambda^0$ transition
  in the Effective quark model with chiral $U(3)\times U(3)$ symmetry.}
\end{figure}
In the coordinate representation the analytical expression
of the matrix element (\ref{label3.1}) takes the form
\begin{eqnarray}\label{label3.6}
  \hspace{-0.3in}&&\langle
  \Lambda^0(k_{\Sigma^0},\sigma_{\Sigma^0})\vert J^{\mu}(0)\vert
  \Sigma^0(k_{\Lambda^0},\sigma_{\Lambda^0})\rangle = -\,\frac{\sqrt{3}}{2}\,g^2_B\int d^4xd^4y
  \,e^{\textstyle\,ik_{\Lambda^0}\cdot x - ik_{\Sigma^0}\cdot y}\nonumber\\
  \hspace{-0.3in}&&\times\, \bar{u}_{\Lambda^0}(k_{\Lambda^0},\sigma_{\Lambda^0}) 
  \gamma^{\alpha}\gamma^5 
  S^{(d)}_F(x - y)\gamma_{\beta} S^{(u)}_F(y) \gamma_{\mu} S^{(u)}_F(-\,x)\gamma_{\alpha}
  S^{(s)}_F(x - y)\gamma^{\beta}\gamma^5 u_{\Sigma^0}(k_{\Sigma^0},\sigma_{\Sigma^0})\nonumber\\
  \hspace{-0.3in}&&- \frac{\sqrt{3}}{2}\,g^2_B\int d^4xd^4y
  \,e^{\textstyle\,ik_{\Lambda^0}\cdot x - ik_{\Sigma^0}\cdot y}\,
{\rm tr}\{\gamma_{\alpha}S^{(s)}_F(x - y)\gamma_{\beta}S^{(d)}_F(y - x)\}\nonumber\\
  \hspace{-0.3in}&&\times\, \bar{u}_{\Lambda^0}(k_{\Lambda^0},\sigma_{\Lambda^0}) 
  \gamma^{\alpha}\gamma^5 
  S^{(u)}_F(x) \gamma_{\mu} S^{(u)}_F( - y)\gamma^{\beta}\gamma^5 
u_{\Sigma^0}(k_{\Sigma^0},\sigma_{\Sigma^0})\nonumber\\
\hspace{-0.3in}&&-\,\frac{\sqrt{3}}{2}\,g^2_B\int d^4xd^4y
  \,e^{\textstyle\,ik_{\Lambda^0}\cdot x - ik_{\Sigma^0}\cdot y}\nonumber\\
  \hspace{-0.3in}&&\times\, \bar{u}_{\Lambda^0}(k_{\Lambda^0},\sigma_{\Lambda^0}) 
  \gamma^{\alpha}\gamma^5 
  S^{(u)}_F(x - y)\gamma_{\beta} S^{(d)}_F(y) \gamma_{\mu} S^{(d)}_F(-\,x)\gamma_{\alpha}
  S^{(s)}_F(x - y)\gamma^{\beta}\gamma^5 u_{\Sigma^0}(k_{\Sigma^0},\sigma_{\Sigma^0})\nonumber\\
  \hspace{-0.3in}&&- \frac{\sqrt{3}}{2}\,g^2_B\int d^4xd^4y
  \,e^{\textstyle\,ik_{\Lambda^0}\cdot x - ik_{\Sigma^0}\cdot y}\,
{\rm tr}\{\gamma_{\alpha}S^{(s)}_F(x - y)\gamma_{\beta}S^{(u)}_F(y - x)\}\nonumber\\
  \hspace{-0.3in}&&\times\, \bar{u}_{\Lambda^0}(k_{\Lambda^0},\sigma_{\Lambda^0}) 
  \gamma^{\alpha}\gamma^5 
  S^{(d)}_F(x) \gamma_{\mu} S^{(d)}_F( - y)\gamma^{\beta}\gamma^5 
u_{\Sigma^0}(k_{\Sigma^0},\sigma_{\Sigma^0}).
\end{eqnarray}
We have taken into account that only the isovector part
$(\bar{u}_{\ell}(0)\gamma_{\mu}u_{\ell}(0) - \bar{d}_{\ell}(0)\gamma_{\mu}d_{\ell}(0))/2$ of the
electromagnetic current gives the contribution to the transition
$\Sigma^0 \to \Lambda^0$.

In the momentum representation Eq.(\ref{label3.6}) reduces to the form
\begin{eqnarray}\label{label3.7}
  \hspace{-0.3in}&&\langle
  \Lambda^0(k_{\Sigma^0},\sigma_{\Sigma^0})\vert J^{\mu}(0)\vert
  \Sigma^0(k_{\Lambda^0},\sigma_{\Lambda^0})\rangle = \frac{\sqrt{3}}{2}\,g^2_B
  \bar{u}_{\Lambda^0}(k_{\Lambda^0},\sigma_{\Lambda^0})\int \frac{d^4k_1}{(2\pi)^4 i} 
  \frac{d^4k_2}{(2\pi)^4 i} \nonumber\\
  \hspace{-0.3in}&&\times \Bigg[
  \gamma^{\alpha}\gamma^5 \frac{1}{m_d - \hat{k}_2}\gamma_{\beta} \frac{1}{m_u - \hat{k}_1}
  \gamma_{\mu} \frac{1}{m_u - \hat{k}_{\Sigma^0} + \hat{k}_{\Lambda^0} - \hat{k}_1}\gamma_{\alpha}
  \frac{1}{m_s - \hat{k}_{\Sigma^0} - \hat{k}_1 + \hat{k}_2}\gamma^{\beta}\gamma^5\nonumber\\
  \hspace{-0.3in}&& +
  {\rm tr}\Big\{\gamma_{\beta}\frac{1}{m_d - \hat{k}_2}\gamma_{\alpha}
  \frac{1}{m_s - \hat{k}_2 - \hat{k}_{\Lambda^0} + \hat{k}_1}\Big\}
  \gamma^{\alpha}\gamma^5 \frac{1}{m_u - \hat{k}_1}
  \gamma_{\mu} \frac{1}{m_u - \hat{k}_{\Sigma^0} + \hat{k}_{\Lambda^0} - \hat{k}_1}
  \gamma^{\beta}\gamma^5\nonumber\\
  \hspace{-0.3in}&&+ 
  \gamma^{\alpha}\gamma^5 \frac{1}{m_u - \hat{k}_2}\gamma_{\beta} \frac{1}{m_d - \hat{k}_1}
  \gamma_{\mu} \frac{1}{m_d - \hat{k}_{\Sigma^0} + \hat{k}_{\Lambda^0} - \hat{k}_1}\gamma_{\alpha}
  \frac{1}{m_s - \hat{k}_{\Sigma^0} - \hat{k}_1 + \hat{k}_2}\gamma^{\beta}\gamma^5\nonumber\\
  \hspace{-0.3in}&& + 
  {\rm tr}\Big\{\gamma_{\beta}\frac{1}{m_u - \hat{k}_2}\gamma_{\alpha}
  \frac{1}{m_s - \hat{k}_2 - \hat{k}_{\Lambda^0} + \hat{k}_1}\Big\}
  \gamma^{\alpha}\gamma^5 \frac{1}{m_d - \hat{k}_1}
  \gamma_{\mu} \frac{1}{m_d - \hat{k}_{\Sigma^0} + \hat{k}_{\Lambda^0} - \hat{k}_1}
  \gamma^{\beta}\gamma^5\Bigg]\nonumber\\
  \hspace{-0.3in}&&\times\,u_{\Sigma^0}(k_{\Sigma^0},\sigma_{\Sigma^0}).
\end{eqnarray}
Applying the heavy--baryon technique, accepted for the analysis of
baryon exchanges within Chiral Perturbation Theory (ChPT) \cite{JG83}
(see also \cite{EW79,AI00} and \cite{TE05}), to the calculation of the
momentum integrals \cite{AI1} we get (see also \cite{IV4}):
\begin{eqnarray}\label{label3.8}
  g_{\Sigma^0\Lambda^0} =  
  \frac{g^2_B}{\sqrt{3}}\,\frac{\langle \bar{q}q\rangle}{m_N}\,
  \frac{m}{16\pi^2}\,{\ell n}\Big(1 + \frac{\Lambda^2_{\chi}}{m^2}\Big) = 
  -\,8.62\times 10^{-4}\,{\rm MeV}^{-4},
\end{eqnarray}
where $\langle \bar{q}q\rangle $ is the quark condensate \cite{AI1,EQM}:
\begin{eqnarray}\label{label3.9}
  \langle \bar{q}q\rangle = -\,\frac{3m}{4\pi^2}\,\Big[\Lambda^2_{\chi} - 
m^2\,{\ell n}\Big(1 + \frac{\Lambda^2_{\chi}}{m^2}\Big)\Big] = (-\,253\,{\rm MeV})^3,
\end{eqnarray}
calculated at the normalization scale $\mu = \Lambda_{\chi}$, where
$\Lambda_{\chi} = 940\,{\rm MeV}$ is the scale of the spontaneous
breaking of chiral symmetry \cite{AI1}, and for $m_u = m_d = m =
330\,{\rm MeV}$, the constituent quark mass defined in the chiral
limit \cite{AI1,EQM}.  For the calculation of the integrals we have
neglected the mass differences between $\Sigma^0$, $\Lambda^0$ and $N$
and have set $m_{\Sigma^0} = m_{\Lambda^0} = m_N$.  Hence, the
accuracy of the theoretical value (\ref{label3.8}) is not better than
$20\,\%$.

For the effective coupling constant (\ref{label3.8}) the transition
magnetic moment $\mu_{\Sigma^0\Lambda^0}$ is equal to
\begin{eqnarray}\label{label3.10}
  \mu_{\Sigma^0\Lambda^0} = 2m_N g_{\Sigma^0\Lambda^0} =  
  \frac{g^2_B}{\sqrt{3}}\,\langle \bar{q}q\rangle\,
  \frac{m}{8\pi^2}\,{\ell n}\Big(1 + \frac{\Lambda^2_{\chi}}{m^2}\Big) = -\,1.62,
\end{eqnarray}
which agrees well with both the experimental data $\vert
\mu^{\exp}_{\Sigma^0\Lambda^0}\vert = (1.61 \pm 0.08)$ and the
theoretical value $\mu_{\Sigma^0\Lambda^0} =
(\sqrt{3}/4)\,(\mu_{\Sigma^-} - \mu_{\Sigma^+}) = (-\,1.57 \pm 0.01)$,
obtained within the naive quark model in terms of the magnetic moments
of the charged $\Sigma$\,--\,hyperons $\mu_{\Sigma^-} = (-\,1.160 \pm
0.025)$ and $\mu_{\Sigma^+} = (2.458 \pm 0.010)$ \cite{PDG04}. In the
naive quark model the magnetic moments $\mu_{\Sigma^-}$ and
$\mu_{\Sigma^+}$ are defined in terms of the magnetic moments of
constituent quarks $\mu_{\Sigma^-} = (4\mu_d - \mu_s)/3$ and
$\mu_{\Sigma^+} = (4\mu_u - \mu_s)/3$. For the transition magnetic
moment $\mu_{\Sigma^0\Lambda^0}$ this gives $\mu_{\Sigma^0\Lambda^0} =
(\mu_d - \mu_u)/\sqrt{3}$ \cite{PDG04a}.

As a result of the calculation of the momentum integrals in
Eq.(\ref{label3.7}) there appears the term proportional to
$\bar{u}_{\Lambda^0}\gamma^{\mu}u_{\Sigma^0}$. Since in the
heavy--baryon limit the spatial part of this current
$\bar{u}_{\Lambda^0}\vec{\gamma} u_{\Sigma^0}$ is of order of
$O(1/m^2_N)$, in the heavy--baryon limit it is equivalent to zero
relative to the contribution proportional to
$\bar{u}_{\Lambda^0}\sigma^{\mu\nu}k_{\nu}u_{\Sigma^0}$.  In turn, the
time--component $\bar{u}_{\Lambda^0}\gamma^0u_{\Sigma^0}$ vanishes due
to the orthogonality of the spinorial functions
$\varphi^{\dagger}_{\Lambda^0}\varphi_{\Sigma^0} = 0$.

\section{Conclusion}

We have analysed the angular distribution of the
$\Lambda^0$\,--\,hyperons, produced in the $\Sigma^0 \to
\Lambda^0\gamma$ decay, both in the laboratory and in the rest frame
of the $\Sigma^0$\,--\,hyperon in dependence on the polarizations of
baryons.  We have calculated the dynamical polarization vector of the
$\Lambda^0$\,--\,hyperon. The obtained results can be used for the
analysis of the polarization properties of the
$\Sigma^0$\,--\,hyperons, produced in high--energy nucleus--nucleus
collisions and in the ${^2_{\!\bar{K}}}{\rm H} \to \Sigma^0 p$ decays
of the Kaonic Nuclear Cluster ${^2_{\!\bar{K}}}{\rm H}$ (the deeply
bound $K^-pp$ state). A measurement of the polarization of the
$\Sigma^0$\,--\,hyperon can be carried out through the measurement of
the polarization of the $\Lambda^0$\,--\,hyperon.

The partial width of the $\Sigma^0 \to \Lambda^0\gamma$ decay is
defined by the transition magnetic moment $\mu_{\Sigma^0\Lambda^0}$.
The calculation of $\mu_{\Sigma^0\Lambda^0}$ we have carried out
within the Effective quark model with chiral $U(3)\times U(3)$
symmetry. The theoretical value $\mu_{\Sigma^0\Lambda^0} = -\,1.62$
agrees well with both the experimental one
$\mu^{\exp}_{\Sigma^0\Lambda^0}\vert = (1.61 \pm 0.08)$ and the
predicted within the naive quark model $\mu_{\Sigma^0\Lambda^0} =
(\sqrt{3}/4)\,(\mu_{\Sigma^-} - \mu_{\Sigma^+}) = (-\,1.57 \pm 0.01)$.

In our approach the transition magnetic moment
$\mu_{\Sigma^0\Lambda^0}$ does not depend on the mass of constituent
$s$\,--\,quark. This agrees well with the naive quark model, where
$\mu_{\Sigma^0\Lambda^0}$ is defined by the magnetic moments of the
constituent $u$\,--\, and $d$\,--\,quarks only:
$\mu_{\Sigma^0\Lambda^0} = (\sqrt{3}/4)\,(\mu_{\Sigma^-} -
\mu_{\Sigma^+}) = (\mu_d - \mu_u)/\sqrt{3}$ \cite{PDG04a}.


\begin{thebibliography}{9}
\bibitem{AI1a}
The coupling constant $g_B$ is defined by \cite{AI1}:
\[g_{\pi NN} = g^2_B\frac{2}{3}\,\frac{m}{F_{\pi}}\,\frac{\langle
  \bar{q}q\rangle^2}{m^2_N},\] where $g_{\pi NN} = 13.21$ is the $\pi
NN$ coupling constant, $F_{\pi} = 92.4\,{\rm MeV}$ is the PCAC
constant of $\pi$\,--\,meson, $\langle \bar{q}q\rangle = (-\,253)^3\,{\rm
  MeV}^3$ is the quark condensate (\ref{label3.9}), and $m = 330\,{\rm
  MeV}$ is the constituent quark mass, calculated in the chiral limit.
This results in $g_B = 1.37\times 10^{-4}\,{\rm MeV}^{-2}$.
\end{thebibliography}
\end{document}